\documentclass[aps,pre,floatfix,nofootinbib,showkeys]{revtex4}
\usepackage{enumerate}
\usepackage{xcolor}
\usepackage{subfigure}
\usepackage{epic}\usepackage{eepic}
\usepackage{bm}
\usepackage{epsfig}
\usepackage{graphicx} 
\graphicspath{ {images/} }
\usepackage{amsmath} \usepackage{amssymb}
\newcommand{\comment}[1]{}

\renewcommand{\P}{\mathfrak{L}}
\newcommand{\F}{\mathfrak{F}}

\renewcommand{\a}{\alpha}
\renewcommand{\b}{\beta}

\newcommand{\be}{\begin{eqnarray}}
\newcommand{\ee}{\end{eqnarray}}

\newcommand{\no}{\label}

\newcommand{\I}{{\rm L}}
\newcommand{\II}{{\rm F}}
\newcommand{\CII}{{\rm CF}}
\newcommand{\TII}{{\rm TF}}

\begin{document} 

\title{Leadership scenarios in  prisoner's dilemma game}

\author{S.G. Babajanyan$^1$, A.V. Melkikh$^2$, and A.E. Allahverdyan$^1$\footnote{armen.allahverdyan@gmail.com}}

\affiliation{$^1$Alikhanyan National Laboratory (Yerevan Physics
  Institute), Alikhanian Brothers Street 2, Yerevan 375036, Armenia}
\affiliation{$^2$Ural Federal University,
  Mira Street 19, Yekaterinburg 620002, Russia}

\begin{abstract} The prisoner's dilemma game is the most known
contribution of game theory into social sciences. Here we describe new
implications of this game for transactional and transformative
leadership. While the autocratic (Stackelberg's) leadership is
inefficient for this game, we discuss a Pareto-optimal scenario, where
the leader $\I$ commits to react probabilistically to pure strategies of
the follower $\II$, which is free to make the first move.  Offering
$\II$ to resolve the dilemma, $\I$ is able to get a larger average
pay-off. The exploitation can be stabilized via repeated interaction of
$\I$ and $\II$, and turns to be more stable than the egalitarian regime,
where the pay-offs of $\I$ and $\II$ are equal.  The total (summary)
pay-off of the exploiting regime is never larger than in the egalitarian
case. We discuss applications of this solution to a soft method of
fighting corruption and to modeling the Machiavellian leadership.
Whenever the defection benefit is large, the optimal strategies of $\II$
are mixed, while the summary pay-off is maximal. One mechanism for
sustaining this solution is that $\I$ recognizes intentions of $\II$.

\end{abstract}

\keywords{leadership; followership; prisoner's dilemma; hierarchic games; 
mixed strategies}
 
\maketitle 

\section{Introduction}

The phenomenon of leadership is studied in various disciplines (biology,
psychology, sociology, management science) and led to a big literature;
see \cite{encyclo,ackoff,traits,autocratic_review,king,jtb,jtb2,evolution,hogan} for
reviews.  Still there is a common opinion that the research on
leadership lacks integrity \cite{hogan,hackman}. One reason for this
is a shortage of transparent mathematical models. 

Existing models divide into two groups. Agent-based models focus on
entities that can modify their interaction due to leaders
\cite{zimm,zimm2,hazy,pais,galam,armen}. In particular, the activity of agents
can be mediated and regulated by leaders \cite{zimm,zimm2,armen}. Such
models can mimic leadership scenarios known in society \cite{armen}, but
the leader-follower interaction assumed by them is frequently
oversimplistic; e.g. it lacks notions of fairness, efficiency {\it etc}. 

Within game-theoretical approaches|both in economic
\cite{king,calvert,colomer} and evolutionary \cite{jtb,jtb2,evolution}
set-ups|the attention of leadership researchers is focused on
Stackelberg's solution in coordination games
\cite{king,calvert,colomer,jtb,evolution}. This provides the most
traditional understanding of leadership, where the leader $\I$ makes the
first move and thereby imposes a solution of the game on the follower
$\II$, which reacts (best-responses) to actions of the leader
\cite{st,zamir,conitzer}. Stackelberg's solution describes certain
leadership scenarios in human communities or in animal groups
\cite{jtb,evolution}. But the leadership phenomenon is richer than
follower(s) reacting on leader's actions.

Leaders are also studies within evolutionary game-theoretic models on
graphs, in particular on complex networks; see
\cite{epjb,biosystems,physrep} for reviews. Here leaders are associated
with strongly connected nodes (i.e. hubs) of the underlying network.
Hubs can influence the behavior of other nodes. 
But relating hubs with leaders is incomplete \cite{aram}, since static networks
do not contain information on the dynamics of influences, e.g. on
information transfer from one node to another \cite{aram}. 

We study the prisoner's dilemma game that has many realizations in economics,
politics and social relations
\cite{luce,shubik,myerson,hofbauer,peterson,unto,sandler,dyson,liu}; see
section \ref{remind} for a reminder.  Here the jointly effective
(cooperative) actions of both player are unstable with respect to a
unilaterial change (defection), which leads both players to inefficient
(jointly defecting), low pay-off state. Yet this state is robust, since
it is a Nash equilibrium of the game. It is also obtained via
Stackelberg's solution, or via the concept of dominant strategies.
Resolving the prisoner's dilemma amounts to finding mechanisms that can
lead to avoiding this inefficient state, thereby going beyond simple
rationality ideas related to the Nash equilibrium or dominant strategies
\cite{shubik,myerson,hofbauer,peterson,unto,sandler,dyson}. 

Our aim here is to work out a class of leadership scenarios that emerge
from within the prisoner's dilemma game. The previous work on leadership in
social dilemma games studied an exogenous leader that initiates
cooperation between several (at least 3) followers engaged in a
prisoner's dilemma-type multi-round game \cite{bates}. In contrast, the present
work will focus on the two-player prisoner's dilemma game, where one of players
shows features of a transactional and transformational leader. Our
interest will be in psychological and game-theoretic mechanisms that
enforce such leadership already in a one-shot (single-round) situation.
Here is a brief description of our model and its main results. 

{\it (i)} We describe a leadership scenario, where $\I$ commits to react
probabilistically to actions of $\II$, which makes the first move; see
section \ref{condo}. The reaction probabilities of $\I$ are determined
from the conditional maximization of the average pay-off of $\I$, which
is conditioned by the fact that the average pay-off of $\II$ is
larger than its guaranteed value.  Under certain conditions this leadership
scenario leads to Pareto-optimal solutions; see section \ref{preto}.
The scenario is similar to transactional leadership \cite{encyclo},
since the response of $\I$ amounts to encouragement|if $\II$ cooperates,
then $\II$ gets a higher average pay-off than the guaranteed value|and
discouragement, where the pay-off of non-cooperating $\II$ is reduced to
its guaranteed value. Also, the situation has basic features of the
transformative leadership \cite{encyclo}: {\it (1.)} $\II$ is free to
make the first move and accept the game rules \cite{ackoff}. {\it (2.)}
$\I$ is able to take $\II$ out of the non-cooperative behavior, i.e. out
of the dominant strategy of $\II$. 

{\it (ii)} The model has two regimes: egalitarian {\it versus}
exploiting leadership. In the first regime $\I$ helps $\II$ to resolve
the prisoner's dilemma, but gets the same {\it average} pay-off as
$\II$. However, {\it generically} $\I$ exploits $\II$, since the average
pay-off of $\I$ is larger. Importantly, this regime is more stable than
the egalitarian one. Note that the leadership scenario is to be
stabilized with respect to $\I$ breaking his commitment (i.e.
deceiving). One|but not the only|stability mechanism comes from a
repeated-game implementation of the scenario, where different rounds of
the game are independent and identically distributed; see section
\ref{stability}. There the stability comes from a back-reaction of $\II$
on $\I$ (i.e. $\II$ punishes $\I$), when $\I$ breaks the commitment.
$\II$ is well-motivated to punish $\I$, since $\II$ looses less out of
this punishment; this mechanism is absent in multi-game implementations
of the usual prisoner's dilemma. Thus the model provides a mechanism
for transition from egalitarian leaders to exploiting ones, a widely
discussed problem in the leadership research
\cite{king,jtb,jtb2,evolution}. 

{\it (iii)} When the pair $\I+\II$ is viable as a whole, i.e.  when
their summary pay-off is large? The summary pay-off can be related to
group fitness \cite{unto} or, alternatively, to the overall resource
extracted from the environment.  Hence this question concerns the
mission of leaders that are supposed to organize efficient, competitive
groups \cite{hogan}. The model shows that with respect to the summary
pay-off, the exploiting regime is never better than a certain
egalitarian one (which can be more complex in its implementation); see
sections \ref{summacontrapaganes} and \ref{preto}. 

{\it (iv)} The exploiting regime assumes a follower $\II$ that agrees
to get less than $\L$, and is supposed to cooperate with $\I$, while
$\I$ will defect $\II$ with a certain probability. Though $\II$ does
have a freedom of defecting, he will be discouraged (punished) by
defection in response. A closer look to the situation
shows that the exploiting leader has features of manipulative
(Machiavellian) personality, as described in experimental literature
\cite{anne,rudinow,dotsenko} \footnote{It is also well-known that animal
leaders manipulate their followers in order to gain more \cite{king}.}.
Hence it can serve as a game-theoretic formalization of the
Machiavellian leadership \cite{mac1,mac2}; see section \ref{disco}. 
How $\II$ can evolve given an exploiting $\I$?
One possibility is that a greedy (pay-off maximizing) $\I$ will lead to an
apathetic $\II$, whose actions have nearly the same average pay-off.  If
the defection benefit is high (i.e. $T>2R$ in (\ref{bad})), $\I$ can
allow $\II$ to employ a mixed strategy, i.e.  to defect with a definite
probability without necessarily punishing him; see section
\ref{tarantul}. This solution is Pareto-optimal for $T>2R$, but it is
also more complex and difficult to manage, because it requires that $\I$
recognizes intentions of $\II$; see section \ref{control}. Once for
$T>2R$ the complexity is managed and the scenario is implemented, $\I$
does not have to exploit $\II$ generically, and also the overall pay-off
of $\I+\II$ is equal to its maximal possible value. 

We organized this paper as follows. Section \ref{remind} reminds the
prisoner's dilemma game and sets notations. Section \ref{condo} presents the
main leadership scenario for resolving the prisoner's dilemma, discusses
its stability, and the summary pay-off. Section \ref{applo} provides a
realistic example of this scenario based on a soft method of fighting
corruption in developing countries. Section \ref{summa} discusses
relation with literature. Section \ref{control} deduces the leadership
scenario from a more general set-up, where $\II$ can act
probabilistically (i.e. within a mixed strategy). Section \ref{control}
also studies the prisoner's dilemma game in the regime, where the defection
benefit is high (i.e. $T>2R$). Section \ref{disco} summarizes our
results in the context of open questions in the leadership research.  A
minimalist reader interested only in basic implication of our results
for leadership theories can read sections \ref{remind}, \ref{condo}, and
\ref{disco}. 

\comment{ Within this leadership scenario, the leader $\I$ can naturally
invest part of his pay-off into motivating the proper behavior of the
follower $\II$.  In particular, $\I$ can apply discouragement instead of
punishment.  We show that this enables cooperative behavior in several
important games, where|within standard solution concepts such as
correlated equilibrium, Stackelberg's equilibrium, dominated strategies
exclusion|the cooperativity is unstable (as in the prisoner's dilemma),
or non-existent as in antagonistic (generalized zero-sum) games.  The
logic of the new solution provides transparent criteria for determining
its stability (commitment credibility). Since the new $\I$ can change
the perspective of $\II$ and provide cooperative solutions where other
leaders failed, we believe it qualifies for a transformative leader.  }

\section{Leadership scenarios for the prisoner's dilemma game}

\subsection{The game and the dilemma: a remainder}
\label{remind}

There are two players, $\I$
and $\II$.  Each one can apply two actions: $c$ (cooperate) and $d$
(defect). Pay-offs are determined by the following matrix
\begin{eqnarray}
\label{bad}
\begin{tabular}{||c|c|c||}
  \hline
~$ {\I}/{\II}$~  & $c$ & $d$ \\
  \hline
  ~$c$~ & ~$R, \, R$~ & ~$S, \, T$~ \\
  \hline\hline
  ~$d$~ & ~$T, \, S$~ & ~$P, \, P$~ \\
  \hline
\end{tabular}\,,
\ee
where e.g. actions $c$ and $d$ by (resp.) $\I$ and $\II$ result to
pay-offs $S$ and $T$ to (resp.) $\I$ and $\II$. The prisoner's dilemma game
is specified by the following relation between the pay-offs:
\be
\no{pri}
T>R>P>S.
\ee
Since all pay-offs can
be simultaneously multiplied by a positive number and added an
arbitrary number, we choose without loss of generality:
\be
\no{pri2}
R=1, \qquad S=0, \qquad {\rm hence}\quad 1>P>0.
\end{eqnarray}
Eqs.~(\ref{bad}, \ref{pri}) show that for both players $d$ is a dominant
strategy, i.e. $d$ yields a higher pay-off than $c$, no matter what the
opponent does. It follows that no player can gain by unilaterally
changing his strategy only if both defect; i.e. $(d,d)$ is the only Nash
equilibrium of game (\ref{bad}). Would both acted $c$, they
would both get $R$, which is larger than $P$ in the Nash equilibrium. But playing
$c$ is vulnerable, since the opponent can change to defecting, gain out
of this, and leave the cooperator with the minimal pay-off $S=0$. This
makes the famous prisoner's dilemma \cite{shubik,myerson,hofbauer,peterson}. 

Eq.~(\ref{bad}) defines a symmetric game: both players are
equivalent on the level of pay-offs and actions. One can try to break
this equivalence on the level of behavior, i.e. making $\I$ a leader,
and $\II$ his follower. Since the game is symmetric, the asymmetry
between $\I$ and $\II$ refers to their attributes, i.e. it is not
imposed by pay-offs (\ref{bad}) of the game. 

One leadership scenario is notoriously unsuccessful in solving the
dilemma. It is based on the concept of Stackelberg's solution
\cite{st,zamir,conitzer}, which is a game-theoretic manifestation of
coercive (or autocratic) leadership \cite{autocratic_review}. Within
this solution, $\I$ makes the first move, knowing that $\II$ will
respond by his best reponse to every move of $\I$. Since the best
response of $\II$ is always $d$, both players will end up in $(d,d)$.
Stackelberg's solution is also inefficient for the sequentially played
prisoner's dilemma game, where it is also known as the backward induction
\cite{myerson}.

\subsection{Solution via conditional probabilistic commitment }
\no{condo}

We propose to solve the dilemma by keeping the idea of leadership, but
now the leader $\I$ does not impose a solution on the follower $\II$.
In contrast, $\II$ is given freedom to decide for himself, while
$\I$ provides $\II$ with probabilities of his reactions to various
actions of $\II$ (i.e. $\I$ commits to a probabilistic reaction
\cite{schelling}): if $\II$ chooses $d$, then $\I$ will take $d$ in
response. If $\II$ will act $c$ (cooperation), then $\I$ will cooperate
with probability $1>{P}+\delta>0$:
\be
\label{de}
{\rm Pr}[\I\mapsto d\,|\, \II\mapsto d\,]=1, \quad
{\rm Pr}[\I\mapsto c\,|\, \II\mapsto c\,]=
{P}+\delta, \quad \delta>0,
\ee
where $\I\mapsto d$ means $\I$ acting $d$, ${\rm Pr}[...\,|\,...]$
means conditional probability, and $\delta$ is a parameter to be 
discussed below. 
If $\II$ acts $c$, then $\I$ and $\II$ get (resp.) the following {\it average} pay-offs:
\be
\label{shan}
&&\P_c=({P}+\delta)\times 1+(1-{P}-\delta)\times T,\\
\no{caroussel}
&&\F_c=P+\delta .
\ee
If $\II$ acts $d$, then both get
\be
\no{an}
\P_d=\F_d=P.
\ee
For $\II$ it is beneficial to cooperate (act $c$), since in average he
gets more than for defecting: $\F_c>\F_d$ due to $\delta>0$. This is why
$\II$ chooses to act $c$. Eq.~(\ref{shan}) shows that $\P_c$ increases
with decreasing $\delta$. Hence a greedy $\I$ would prefer to make
$\delta$ smaller, but for $\delta\to 0$ the situation becomes
meaningless, since $c$ and $d$ are of equal value for $\II$. 

According to (\ref{shan}, \ref{caroussel}) there is a clear asymmetry
between $\I$ and $\II$: $\I$ gets (in average) larger than $R$,
$\P_c>R=1$, while $\II$ gets smaller than $R$, $\F_c<R=1$, but still
larger than for the defecting strategy ($\F_d=P$). Below we shall see
that this asymmetry can make the situation more stable.
Note that for following these rules of the game, $\I$ has to known all
pay-off of the game, while $\II$ may|but need not|know pay-offs of $\I$. 
For simplicity and definiteness we shall assume that $\II$
knows the pay-offs of $\I$. 

Section \ref{control} shows that (\ref{de}--\ref{an}) come out from an
optimization principle, where $\I$ tries to maximize his {\it
conditional average} pay-off; i.e. the average pay-off given that $\II$
acts $c$ and gets an average pay-off larger than $P$. Also, provided
$T<2R=2$, solution (\ref{de}--\ref{an}) is Pareto-optimal, i.e. it is
impossible to increase the average pay-offs of {\it both} $\I$ and $\II$
by deviating from it; see section \ref{control}. We emphasize that
although we are looking at a single-shot game, the concept of average
pay-off does apply; see e.g. \cite{myerson}. 

\comment{ Moreover, one can show that the prisoner's dilemma game is the only
symmetric game that admits such a probabilistic response of $\L$ within
the above Nash equilibrium \cite{sano}.  For other symmetric games the
maximization of the conditional average reduces to taking a pure
strategy in response to $\II$ \cite{sano}.  }

What will prevent $\I$ from deceiving, i.e. announcing probabilistic
response via (\ref{de}), but acting $d$ (i.e. his best response) with
probability one?  One answer to this question relates to the repeated
game implementation of (\ref{shan}--\ref{an}) and is discussed below.
Another possible answer is that the leader $\I$ will have some prestige
losses if his initial promise is not followed. Then the solution is
stable if such losses are larger than $T-\P_c$. Hence the asymmetry
between $\I$ and $\II$ can make the situation more stable, because
$T-\P_c$ can be sufficiently small. Note that the same story on prestige
can be told for the ordinary prisoner's dilemma; cf.~section
\ref{remind}. But there it amounts to a stronger assumption, since each
player should be assumed to have prestige (or reputation) losses larger
than $T-R$, and the summary prestige loss should be larger than
$2(T-R)$. Now $2(T-R)$ is larger (and can be much larger) than $T-\P_c$.
Moreover, introducing prestige losses for the ordinary prisoner's
dilemma amounts to a trivial change of pay-offs in (\ref{bad}) that
amounts to making the outcome $(c,c)$ the Nash equilibrium of the game.

Notions of reputation and prestige became recently popular for
explaining aspects of human cooperative behavior \cite{physrep}. The
relevance of these notions is supported experimentally, e.g. it is known
that people whose behavior is observed tend to cheat less
\cite{physrep}. Ref.~\cite{evol_com} provides a fuller account of
various prestige (reputation) factors; see also
\cite{chin_reputation,nature} for examples and \cite{physrep,gossip,gossip_r} for
reviews.

\subsection{Implementation via repeated games}
\no{stability}

\subsubsection{Stability against deception}

So far we focused on the single-shot game described by
(\ref{bad}--\ref{an}). Recall that according to the law of large
numbers, the averages in (\ref{shan}, \ref{caroussel}) can be realized
as actual pay-offs in sufficiently many identical, independent rounds of
the game.  We assume that the action of $\I$|in particular, if this
action is a deception, i.e. acting $d$, despite of the fact that the
probabilistic mechanism governing (\ref{de}) generated $c$|becomes known
to $\II$ before the next round of the game.  More realistically, $\II$
will have to learn about deceptions of $\I$ via looking at a
sufficiently long statistics of actions of $\I$. 

Importantly, we assume that $\I$ and $\II$ will play $N=N_1+N_2$ times,
where $N\gg 1$ is a sufficiently large number that is not known to $\I$
and to $\II$ \footnote{In particular, even immediately before playing
the last round of the game, $\I$ does not know that it will be the last
play. This assumption is not be met in many real-life examples of
repeated games.  However, it is inavoidable, since without this
assumption the situation is vulnerable to the backward induction, which
essentially trivializes the situation for the prisoner's dilemma game
\cite{myerson}. Indeed, knowing that a given round is the last one, $\I$
(which moves the last one) will certainly act his best-response against
any move by $\II$. Now $\II$ will certainly know this, so he will know
what to play in the last round {\it etc}. So without the above
assumption (and if other assumptions are not made), we shall always be
confined by the situation, where both $\I$ and $\II$ act $d$, i.e. the
prisoner's dilemma holds.}.

\begin{figure}[!ht]
\centering
{\includegraphics[width=8cm]{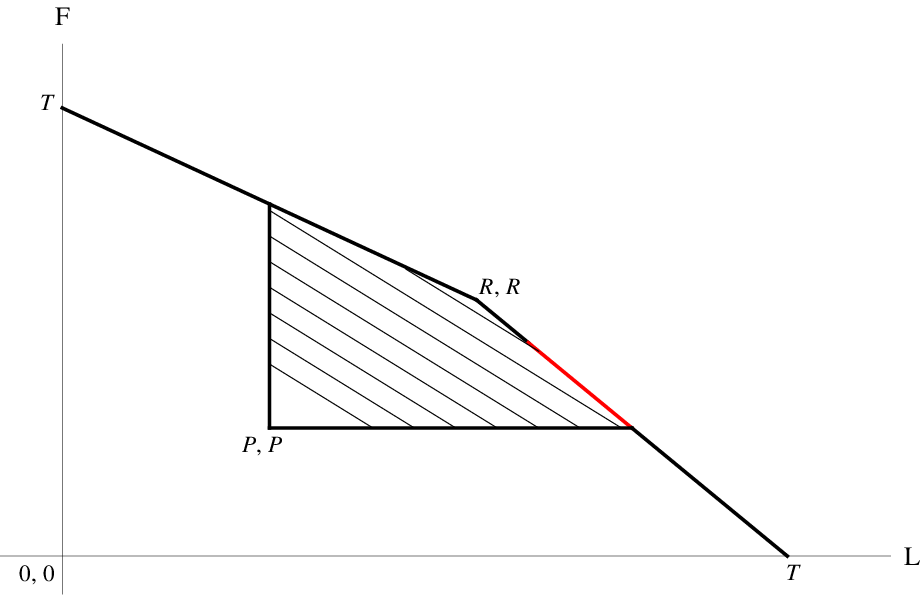}}
\caption{Pay-off diagram of prisoner's dilemma game under conditional probabilistic commitment solution; see section \ref{condo}.
Pay-offs of $\I$ ($x$-axes) and $\II$ ($y$-axes) for prisoner's dilemma parameters $T=\frac{7}{4},
R=1, P=\frac{1}{2}$ in (\ref{bad}, \ref{pri}). At these particular
values of parameters, $\P_c$ and $\F_c$ are given by the red line; the end-points of this
line are $\delta=0$ and $\delta=\frac{T-P}{2(T-1)}-P$. The later value of $\delta$ saturates 
(\ref{somoni}) at $\chi=1$. It is seen that the summary pay-off $\P_c+\F_c$ is smaller than $2R=2$.\\
The stroked region represent individually
rational and feasible pay-offs that appear in the folk theorem for sequentially played (multi-round) game, where 
each round is governed by (\ref{bad}). It is seen that the red line is a part of the 
Pareto front.}
\label{fig2}
\end{figure}

Due to $N\gg 1$, the pay-off accumulated by $\I$ (resp. by $\II$) is
approximately equal to average pay-offs times $N$. We assume that after
each deception, $\II$ opposes $\I$ (punishes $\I$) by acting $\chi$ times $d$. Let us
assume that there are $N_1$ cases, where $\II$ had to respond by $c$,
but acted $d$ (i.e. deceived his commitment). If $\I$ deceived $N_1$
times, he will be opposed $N_2=\chi N_1$ times by $\II$.  This is
compared with a case, where $\I$ never deceives, and the game goes on
$N$ times, i.e. the gain of $\I$ ($\II$) is $N\P_c$ ($N\F_c$);
cf.~(\ref{shan}, \ref{caroussel}). Now if
\be 
\label{somoni} 
N_1T+N_2P<(N_1+N_2)\P_c \qquad {\rm or}\qquad T+P\chi<(1+\chi)\P_c,
\ee 
then deceiving is not beneficial. For a sufficiently large $\chi$,
(\ref{somoni}) always holds due to $\P_c>R=1>P$ in (\ref{shan}). 
If $\II$ never opposes, $\chi=0$, then (\ref{somoni})
reduces to $\P_c>T$ which can never be satisfied, since the average
pay-off cannot be larger than the maximal one. Note that (\ref{somoni}) can be also
satisfied for $\chi<1$, which can be interperted by saying that $\II$ 
opposes with probability $\chi$.

\comment{
When $\II$ punishes only once, $p=1$, for each deception, (\ref{somoni})
reduces to a non-trivial condition on the pay-offs (\ref{bad}) and
probability (\ref{de})
\begin{eqnarray}
\label{pr:2}
  \delta<\frac{1}{2}\,\,\frac{T-P}{T-R}-\frac{P}{R},
\end{eqnarray}
which can hold if the right-hand side of (\ref{pr:2}) is positive, i.e. for
\begin{eqnarray}
  \label{pr:3}
  R>\frac{2P}{1+P/T}.
\end{eqnarray}
If (\ref{pr:3}) holds, then one can choose $\delta$ such that (\ref{pr:2})
holds and the solution is stable. Eq.~(\ref{pri}) shows that
$\frac{2P}{1+P/T}>P$; hence (\ref{pr:3}) is a non-trivial condition. }

It is seen from (\ref{somoni}) that, if the influence of $\II$ on $\I$
is limited, i.e.  when $\chi$ cannot be sufficiently large, there are
situations where the fair strategy (with $\delta=1-P$) is not stable,
while the unfair strategy with $\delta<1-P$ is stable. In such
situations, $\II$ may prefer to accept the unfair rules with a partial
defection of $\I$, because this makes the situation more stable with
respect to deception.

We emphasize an important aspect of above opposing (punishing) by $\F$:
it is well-motivated specifically due to the asymmetry between $\I$ and
$\II$: whenever $\II$ opposes $\I$, $\II$ knows that he will loose
$\F_c-P$, which is smaller than what $\I$ will loose, i.e. $\P_c-P$
\footnote{This mechanism of motivating punishment was described in
Ref.~\cite{luce} (section 5.10), but to our knowledge it was not applied
to prisoner's dilemma.}.  Note that the punishment for the usual
(symmetric) multi-round prisoner's dilemma game does lack such a
motivation. Here the punishing player looses as much as the punished
one. 

\subsubsection{Stability with respect to sequential defection}
\no{seq}

For the sequentially played prisoner's dilemma game it is frequently emphasized
that the situation is non-trivial only under an additional assumption
$2R=2>{T}$ in (\ref{pri}) \cite{hofbauer}. Otherwise, there is an
additional cooperation scenario that shows itself in an even number of
game rounds: the two players can defect each other consequently, i.e. to
act $(c,d)$ in the first tour and $(d,c)$ in the second tour. This leads
to sharing the pay-offs. In this way each one gets $\frac{T}{2}$ per
round, and for $2R=2<{T}$ this is larger than $R$ obtained for $(c,c)$. 

For our situation the condition $1>\frac{T}{2}$ is not critical in the
sense that even for $1<\frac{T}{2}$ the average pay-off (\ref{shan}) of
$\I$ need not be smaller than $\frac{T}{2}$. Hence $\I$ is not
interested to change the situation and the probabilistic solution is
stable with respect to playing $(c,d)+(d,c)$. Indeed, we note that from
(\ref{shan}) $\P_c>\frac{T}{2}$ leads to an upper bound on $\delta$:
\be
\delta<\frac{P+T(\frac{1}{2} -{P})}{T-1}.
\no{krug}
\ee
For $T>2$ the RHS of (\ref{krug}) is smaller than $1-{P}$;
cf.~(\ref{de}). Hence $T>2$ and (\ref{krug}) impose a non-trivial upper
bound on $\delta$ that can be satisfied for a sufficiently small
$\delta$. 

\subsection{Summary pay-off}
\no{summacontrapaganes}

One can look at this situation from a global viewpoint, i.e.
comparing the two leadership strategies|fair vs. exploiting|with
respect to the summary pay-off. Hence we compare $\P_c+\F_c$ from
(\ref{shan}, \ref{caroussel}) with $2R=2$, which is the summary pay-off
for the fair situation.  One motivation for doing this comes from the
group selection ideas (in biology, ecology and economics) \cite{unto},
where the summary pay-off may play the role of fitness if the dyadic
group ($\I+\II$) is looked up as a whole and compared with other groups. 

We get from (\ref{shan}, \ref{caroussel}) 
\be
\P_c+\F_c-2=(T-2)(1-P-\delta),
\no{luka}
\ee
i.e. for $T<2=2R$, the exploiting leadership has a smaller
summary pay-off than for the cooperative strategy $(c,c)$.  Now for $T>2$, we get $\P_c+\F_c>2$, but for $T>2$
there is (for even number of games) another fair and symmetric strategy, {\it viz.}
$(c,d)+(d,c)$ that achieves the summary pay-off equal to $T$, as we saw above. 
Then we have
\be
\no{emu}
\P_c+\F_c-T=-(T-2)(P+\delta),
\ee
which is again negative for $T>2$.  We conclude from (\ref{emu}) that,
at any rate, there is always a fair solution that outperforms the
considered probabilistic solution in terms of the summary pay-off. 

\comment{ For $T>2$ the summary pay-off can be larger than $2R=2$, and a
manipulative $\I$ can employ this fact for explaining that the global
situation is not so bad, even compared with $(c,d)+(d,c)$, where a
higher pay-off is achieved due to additional sychronization between the
actions. }

\subsection{Application: fighting corruption softly }
\no{applo}

We describe here a new applications of the prisoner's dilemma game (\ref{bad},
\ref{pri}), where the solution (\ref{shan}, \ref{caroussel}) is
realistic.  Let $\I$ be a potentially corrupt official (bribee) with
actions $c$ and $d$ referring to, respectively, not taking and taking
bribes. $\II$ is the government that decided to eradicate corruption via
raising salaries of officials by paying them wage bonus. Such
anti-corruption policies were implemented in certain developing
countries, e.g. in Syria during early 1980's \cite{macrae} and in
Armenia more recently \cite{grigoryan}.  There is a clear empiric evidence
across various developing countries showing that low wages can indeed
promote corruption \cite{imf}. Hence $\II$ have to actions: to keep the
wage bonus ($c$) or to skip it ($d$). From the viewpoint of $\I$ in
(\ref{bad}): $P$ is the bribe money, $R=1$ is the bonus and $T=R+P<2R$.
The pay-offs of $\II$ in (\ref{bad}) are interpreted in a similar way,
but they are more convoluted, since they involve both money and
consequences of corruption. Hence the game need not be symmetric with
respect to $\I$ and $\II$, but for clarity we stick to the symmetric
situation (\ref{bad}), also because consequences of corruption are
difficult to estimate quantitatively. 

Now (\ref{shan}, \ref{caroussel}) refers to the situation, where the
government $\II$ does pay the bonus, but $\I$ will still take bribes
with a certain probability. Note that (\ref{luka}) applies due to
$T=R+P<2R$, showing that the summary pay-off is smaller than $2R=2$, i.e.
this method is not efficient from a global viewpoint. To our
understanding the solution described by (\ref{de}, \ref{shan},
\ref{caroussel}) was qualitatively realized in Armenia: extra wages to
officials were never withdrawn (i.e. $\II$ always acts $c$), and the
corruption was controlled but not eradicated; e.g. the country's
transparency index is still low and did not improve \cite{ti}. 

Note that when officials overdo with bribing (i.e. they raise the
probability of defection), the government prosecutes them legally. At
least, this was the main scenario of government actions in Armenia.
Hence within this realization of the prisoner's dilemma game and its
probabilistic commitment solution [cf.~section \ref{condo}], the
punishment of the deceiving leader involves|unlike the repeated
implementation of the prisoner's dilemma game discussed in section
\ref{stability}|actions that go essentially out of the prisoner's
dilemma game itself. 

\section{Relations with literature}
\label{summa}

\subsection{General remarks}

The solution (\ref{shan}, \ref{caroussel}) does have several important
predecessors. To our knowledge, the first example of conditional
probabilistic commitment was mentioned by Schelling \cite{schelling},
who discussed it in the context of dominated strategies of $\I$; see
\cite{evol_com} a recent broad discussion of various aspects of
commitment. Yet another related approach was proposed by Brams
\cite{brams}, who studied the prisoner's dilemma in the context of the
Newcomb's paradox; see also \cite{brams2} in this context. The
probabilistic prediction scheme developed in \cite{brams,brams2} can be
regarded as a particular case of the conditional probabilistic
commitment studied below. The solution (\ref{shan}, \ref{caroussel})
also shares certain features of adaptive strategies found in
\cite{dyson,liu,akin}; see also \cite{gossip} for discussion. However,
these solutions are completely symmetric with respect to players. In
particular, both of them act simultaneously. For those solutions $T<2R$
is a critical condition, in contrast to our situation; see section
\ref{seq}. 

The fair (deterministic response) solution (\ref{de}) coincides with the
one studied by Howard \cite{howard}, within the so called inverse
Stackelberg's solution; see \cite{olsder,groot} for recent reviews. Note
that Howard also proposed the so called second-order metagame, whose
application to the prisoner's dilemma game lead to non-trivial predictions and
was considered to be a solution of the prisoner's dilemma \cite{shubik}.
However, the general approach of the second-order metagames is
convoluted and involves hidden assumptions; see Appendix \ref{meta2} for
a short discussion. 

\subsection{Relations with the ultimatum game}
\label{ulti}

Solution (\ref{de}--\ref{caroussel}) has certain similarities with the
ultimatum game; see e.g. \cite{ulti,skyrms} for in-depth description of
this game. We shall discuss these similarities with two purposes in
mind: first the ultimatum game can provide some information on the
partially constrained parameter $\delta$ in (\ref{de}--\ref{caroussel}).
Second, the ultimatum game itself gives a clear example, where people do
not employ their best-response action. 

Within the ultimatum game, the first player $I$ proposes to the second
player $II$ to divide a certain amount of money $\mathfrak{M}$ according
to two different strategies.  The fair strategy $f$ gives
$\mathfrak{M}/2$ to each player. The second strategy $u$ gives
$0.01\,\mathfrak{M}$ to $II$, while $I$ gets $0.99\,\mathfrak{M}$.  Now
$II$ can just accept the proposal by $I$ (action $a$) or reject it
(action $r$). In the latter case, both $I$ and $II$ get nothing.  Since
$II$ reacts on actions of $I$, the strategies of $II$ are $aa$ (always
accept), $rr$ (always reject), $ar$ (react $a$ on $f$ and $r$ on $u$),
and $ra$. Using (\ref{ultii}) from Appendix \ref{cons}, one can show
that the pairs of strategies $(f,ar)$ and $(u, aa)$ are Nash equilibria,
but only $(u, aa)$ is the subgame-perfect Nash equilibrium, since (ever)
playing $r$ is against the interests of $II$: better a puny pay-off than
nothing \cite{ulti,skyrms}. The concept of sub-game perfectness
postulates that if actually playing $r$ is not rational, then also
treatening to play $r$ is not rational. This postulate is (expectedly)
invalidated in experiments with human subjects playing the ultimatum
game, which routinely act $r$ in response to $u$ \cite{ulti,skyrms}. The
pair $(u,aa)$ is seen experimentally only if the fraction $0.99$ is
replaced by number closer to (but still smaller than) $0.5$; e.g. $0.3$
\cite{ulti,skyrms}. 

The analogy of this situation with (\ref{de}--\ref{caroussel}) is that
there as well $\I$ proposes to $\II$ a deal that is asymmetic (i.e. $\I$
generically benefits more than $\I$), but does neverthelss provide
to $\II$ an average pay-off $P+\delta$ larger than its guaranteed value $P$. Thus
by analogy with experiments on the ultimatum game, we expect that human
followers $\II$ will not accept small values of $\delta$, instead
choosing to act $d$.  Note that $d$ is the dominant strategy for $\II$,
which is not the case with action $r$ for the second player in the
ultimatum game. Moreover, even if $\II$ agrees with the deal, he may
still be deceived by $\I$, who acts the last one.  These arguments
strengthen the point against small values of $\delta$. Another
difference is that in the ultimatum case the summary pay-off is always
fixed and equals its maximal value $\mathfrak{M}$, in contrast to
(\ref{de}--\ref{caroussel}), where the summary pay-off is generically
smaller than $2R$; cf.~(\ref{luka}). 

\comment{
It is natural to ask whether the stability consideration around
(\ref{somoni}) (section \ref{stability}) relates to more traditional
stability notions of repeated game theory \cite{myerson}. We have shown
\cite{sano} that (\ref{somoni}) can be related to the notion of sub-game
perfect Nash equilibrium in the following sense: one can design
multiple-game strategies that have the same pay-off (per-game) as
(\ref{shan}, \ref{caroussel}), and whose conditions for sub-game Nash
perfectness relate to (\ref{somoni}). We prefer the analysis of section
\ref{stability}, because it is clearer and uses less assumptions. The
factor $\chi$ in (\ref{somoni}) can be related to discount factors in
repeated games \cite{sano}. We stress that what we study are {\it not}
standard repeated games, where at each round players act simultaneously
\cite{myerson}. In our situation, $\I$ acts after $\II$ in each round,
i.e. we do have a leadership game. This remark also concerns the
solution (\ref{de}) that superficially resembles (but is different from)
the known tit-for-tat strategy in simultaneous repeated games
\cite{hofbauer}. In particular, the the standard tit-for-tat strategy
has different conditions for stability (sub-game Nash perfectness). }

\subsection{Relations with folk theorems for repeated games}

General features of our implementation of the probabilistic commitment
solution via repeated games (see section \ref{stability}) are governed
by the folk theorem \cite{myerson,aumann,fud}. Hence we shall review it
and specify its relation with our results. We shall focus on the
implementation of repeated games that is similar to what we discussed in
section \ref{stability}. Here the average pay-offs are just arithmetic
averages of single-round pay-offs, which|due to the law of large numbers
and for sufficiently long series of rounds| coincides with probabilistic
averages seen within the single-shot realization of the game \footnote{Note that
there is another mechanism for implementing repeated games. It goes by
introducing the discount factor, which is a finite probability
by which every round of the repeated game takes place. This factor
allows to get finite average pay-offs for a long series of rounds
\cite{myerson,aumann,fud}. We shall not employ this implementation,
because the discount factor (being an additional parameter) makes
pay-offs essentially different from their single-shot analogues.}. Then
the main (folk) theorem of the set-up starts by looking at the convex
sum of individual pay-offs; e.g. for the prisoner's dilemma game (\ref{bad}) these
are vectors of the form
\be
\label{sho}
\alpha_1\times (R,R)+ \alpha_2\times(S,T)+\alpha_3\times(T,S)+\alpha_4\times(P,P),\\
\label{p}
{\sum}_{k=1}^4\alpha_k=1,\qquad \alpha_k\geq 0,\qquad k=1,...,4,
\ee
where $\alpha_1$ is the probability by which the action $(R,R)$ is taken {\it etc}. 
The first (second) component $\alpha_1R+ \alpha_2 S+\alpha_3 T+\alpha_4 P$
($\alpha_1R+ \alpha_2 T+\alpha_3 S+\alpha_4 P$)
of vector (\ref{sho}) refers to the pay-off
by $\I$ ($\II$).  The convex domain defined via (\ref{sho}) is to be
bound by restricting both components of (\ref{sho}) to values larger or
equal than $P$; see the stroked domain in Fig.~\ref{fig2} (for $2R=2>T$) 
and in Fig.~\ref{fig3} (for $2R=2<T$). Now the folk
theorem states that every point within the (stroked) domain can be realized
via a certain joint strategy $\I$ and $\II$ in the repeated game as a
a sub-game perfect Nash equilibrium \cite{aumann}.  This strategy (which
is generally not unique) amounts to to $\I$ and $\II$ acting $(c,c)$ for
the first $n_1$ rounds, then acting $(c,d)$ for subsequent $n_2$ rounds,
acting $(d,c)$ for further $n_3$ rounds, and $(d,d)$ for $n_4$ rounds.  Then the
cycle is repeated, i.e. $(c,c)$ for
$n_1$ rounds {\it etc}. Here $N=M(n_1+n_2+n_3+n_4)\gg 1$ ($M\gg 1$) is the overall number
of rounds, and the relation with the average pay-off vector (\ref{sho})
is obvious: $\alpha_k=\frac{n_k}{n_1+n_2+n_3+n_4}$. If one of players
deviates from this agreement, another one punishes him by acting $d$ for
the rest of the multi-round game. This is the Nash equilibrium, since
none of players will benefit by deviating from this strategy
unilaterially. It is sub-game perfect, since it still decribes a Nash
equilibrium if the game starts from $i$th round ($i>1$, but $i\ll
N$) instead of the first round \cite{aumann}.  

Note that the above construction was realized as deterministic game for
each round, and we also did not specify the order by which $\I$ and
$\II$ act within each round. In contrast, within the implementation in
section \ref{stability}, $\I$ acted probabilistically after $\II$. These
differences are not essential, as far as the above folk theorem
and its implications are concerned. This is seen from the fact that
(\ref{sho}) describes an averaged pay-off. 

Hence according to the folk theorem, all points within the stroked
domain in Figs.~\ref{fig2} and \ref{fig3} have equal status. The folk
theorem is clearly silent about additional mechanisms that are needed
for selecting any specific point or sub-domain there. This is a weak
point of the folk theorem \cite{aumann}. This point gets worst by noting
that the stroked domain is not even restricted to Pareto efficient
pay-offs for $\I$ and $\II$. In contrast, our solution|which is also
realized by sub-game perfect Nash equilibrium strategies, as we saw in
section \ref{stability}|restricts within the stroked domain a sub-set of
Pareto efficient pay-offs; see the red line in Fig.~\ref{fig2} and the
blue line in Fig.~\ref{fig2}. The upper end point of this line is
determined by parameters $\delta$ and $\chi$; see section
\ref{stability}.

\section{General case}
\no{control}\no{preto}

\subsection{Definition}

So far we postulated (\ref{de}) and studied its consequences. Now
relations in (\ref{de}) will be deduced from a general setting, where
$\II$ can employ mixed strategies in his first move. We shall then see
that (\ref{de}) emerges as a Pareto-optimal solution for $T<2R$;
cf.~(\ref{bad}). 

Now $\II$ makes the first move (act) by choosing between two mixed
strategies. Within the first strategy, $\II$ acts $c$ with probability
$x$ and is responded by $\I$ with conditional probabilities $p_1$ and $p_2$:
\be
p_1={\rm Pr}[\I\mapsto c\,|\, \II\mapsto c\,], \quad
p_2={\rm Pr}[\I\mapsto c\,|\, \II\mapsto d\,], \quad x={\rm Pr}[\, \II\mapsto c\,].
\no{dud}
\ee
Within the first strategy, $\II$ acts $c$ with probability
$y$ and is responded by $\I$ with conditional probabilities $q_1$ and $q_2$:
\be
q_1=\overline{{\rm Pr}}[\I\mapsto c\,|\, \II\mapsto c\,], \quad
q_2=\overline{{\rm Pr}}[\I\mapsto c\,|\, \II\mapsto d\,], \quad y=\overline{{\rm Pr}}[\, \II\mapsto c\,].
\no{dudinka}
\ee
This set-up is more complex than the previous case, where $x=1$ and
$y=0$. For now $\I$ has to distinguish from which strategy the defection
of $\II$ came. E.g. this can be done via controlling the probabilistic
mechanism of actions of $\II$. Another possibility is that $\I$ is aware
of intentions of $\II$ when acting $d$, i.e. $\I$ has side (e.g.
emotional) reasons to believe whether $d$ was really acted within the
mixed strategy \footnote{For recent discussion on intentions in game
theory see \cite{intention1,intention2}.}. 

But this more complex set-up has its benefits, because|as seen below in
section \ref{tarantul}|it can improve the pay-off for both $\I$ and
$\II$, for $T>2R$. If $\I$ is not ready to bear this complexity, he will
likely oppose any non-deterministic strategy of $\II$ by treating all
defections in the same way. Such a leader will stick to solution (\ref{de}).

\subsection{Derivation}

$\I$ determines $p_1$, $p_2$, $q_1$, $q_2$, and $y$ in (\ref{dud}) from maximizing his average pay-off
conditioned upon the fact that $\II$ gets more within (\ref{dud}) than within (\ref{dudinka})
[cf.~(\ref{bad}, \ref{pri})]:
\be
\no{aa1}
\P={\rm max}_{\,0\leq p_1,p_2,q_1,a_2,y\leq 1}[~p_1x\times 1+p_2(1-x)\times 0+(1-p_1)x\times T+(1-p_2)(1-x)\times P,  \\
\no{aa11}
p_1x\times 1+p_2(1-x)\times T +(1-p_1)x\times 0+(1-p_2)(1-x)\times P> \\
q_1y\times 1+q_2(1-y)\times T +(1-q_1)y\times 0+(1-q_2)(1-y)\times P
],
\no{aa2}
\ee
where the maximization over $x$ will be caried out later on (for 
the time being it is interesting to leave it as a free parameter). Note that the average pay-off of $\I$,
given by the right-hand-side of (\ref{aa1}), does not depend on $q_1$, $q_2$ and $y$. Hence the maximization
of (\ref{aa1}) is achieved whenever the restriction (\ref{aa2}) is possibly weak, i.e. 
$q_1y +q_2(1-y)\times T +(1-q_2)(1-y) P$ is to be minimized. But the latter quantity|which is 
the average pay-off of $\II$ within (\ref{dudinka})|cannot be smaller than the guaranteed pay-off $P$ of $\II$.
Otherwise, (\ref{dudinka}) is meaningless for $\II$. Hence we should have $q_1y +q_2(1-y)\times T +(1-q_2)(1-y) P=P$,
which is achieved for
\be
q_1=0,\quad q_2=0, \quad y=0,
\ee
i.e. the optimal|from the viewpoint of $\I$|choice of parameters in 
(\ref{dudinka}) amounts to $\II$ defecting with probability $1$, 
achieving the guaranteed pay-off, and then defected 
by $\I$ in response. 

Thus (\ref{aa1}, \ref{aa2}) amount to 
\be
\no{a1}
\P={\rm max}_{\,0\leq p_1,p_2\leq 1}[~p_1x +(1-p_1)x T+(1-p_2)(1-x) P,  \\
\no{a2}
p_1x +p_2(1-x) T +(1-p_2)(1-x) P> P~].
\ee
It is now clear that assumptions (\ref{dud}, \ref{dudinka}) on just two different strategies is not 
essential: any number of them will lead to (\ref{a1}, \ref{a2}) via the same mechanism. 

Constraint (\ref{a2}) can be written as
\be
\label{ukht}
\Delta\equiv p_1x+p_2(1-x)(T-P)-xP>0,
\ee
where $\Delta$ is bounded from above due to $1\geq p_1$ and $1\geq p_2$:
\be 
 \Delta\leq  (1-P)x+(1-x)(T-P).
\no{cau}
\ee
It appears that the straightforward maximization in (\ref{a1}, \ref{a2}) leads
to $\Delta=0$. This is meaningless, since then $\II$ will not be
interested to do anything but defection. Hence we shall take
$\Delta>0$ as a parameter of the solution, express $p_1$ through
$\Delta$ and compute (\ref{a1}) by maximizing over a single parameter $p_2$:
\be
\label{b1}
&& \P=xT(1-P)+P-\Delta (T-1)+T(T-P-1)(1-x)p_2^*,\\
\label{b3}
&& p_2^*=\vartheta[\,T-1-P\,]\,{\rm min}\left[
\frac{\Delta+xP}{(1-x)(T-P)},\, 1
\right],\\
\label{b2}
&& \F=\Delta+P,
\ee
where $\vartheta[x]$ is the step function: $\vartheta[x<0]=0$ and
$\vartheta[x>0]=1$. Eq.~(\ref{b2}) for the average pay-off $\F$ of $\II$
is obtained directly from (\ref{ukht}), where $p_2^*$ is the
optimal value of $p_2$. The corresponding optimal value of $p_1^*$ is
found from (\ref{ukht}). Thus $\Delta$ is a parameter that
governs the difference between the actualy pay-off of $\II$ and its 
guaranteed value; see (\ref{b2}). For $x=1$, we have $\Delta=\delta$ and (\ref{ukht}, \ref{b1}, \ref{b2})
revert to (resp.) (\ref{de}, \ref{shan}, \ref{caroussel}). 

One way of dealing with (\ref{b1}, \ref{b3}) is to maximize $\P(x)$ over
$x$ keeping $\Delta$ fixed, i.e. keeping the average pay-off of $\II$
fixed. To this end, we look at $\P(x)$ for $x\lesssim 1$.  This shows
that for $T<2=2R$, function $\P(x)$ locally maximizes at $x=1$, which is
clearly also the global maximum of $\P(x)$. Putting $x=1$ into
(\ref{a2}--\ref{b2}) we revert to the previous situation; see
(\ref{de}--\ref{caroussel}). It is clear from the structure of
(\ref{a1}, \ref{a2}) that after maximization over $x$ we produce a
Pareto-optimal solution: $\P_c$ and $\F_c$ in (\ref{shan},
\ref{caroussel}) are such that neither can be increased by not
decreasing another; see Fig.~\ref{fig2} for illustration \footnote{More
formally, we get a Pareto-optimal solution by keeping one (average)
pay-off fixed and then maximizing over another \cite{karlin}. This is
because both pay-offs are concave (moreover linear) varying over a
convex set \cite{karlin}.}. 

\subsection{The case $T>2R$}
\label{tarantul}

\begin{figure*}[t]
    \centering
\subfigure[]{
     \label{fig}
\includegraphics[width=7cm]{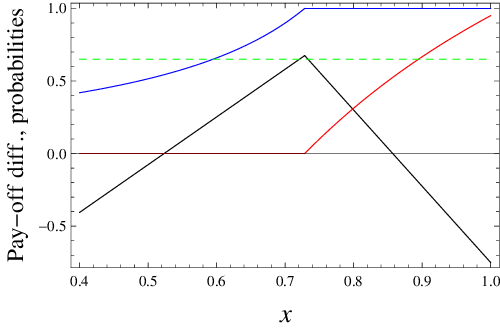}
    }\,\,\,\,\,
\subfigure[]{
     \label{fig1}
\includegraphics[width=7cm]{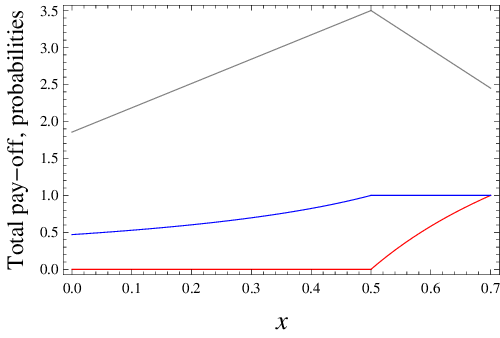}
    }
\caption{These figures illustrate (\ref{b1}--\ref{b2}) for the prisoner's dilemma with $T>2R$ and
compare them with the situation $T<2R$; see (\ref{bad}, \ref{pri}). \\
(a) The pay-off difference $\P-\P_c$ (black curve) given by (\ref{b2}, \ref{shan})
versus $x$ for the prisoner's dilemma game (\ref{bad}, \ref{pri}) with $T=3.5$, $R=1$,
$P=0.2$, $\Delta=0.75$ and $\delta= 0.45$; cf.~(\ref{dud}). Red and blue curves: the
probabilities $p_1^*$ and $p_2^*$ (resp.) given by (\ref{dud}, \ref{ukht}
\ref{b3}). Dashed green line: $P+\delta$; cf.~(\ref{de}). With these
parameters we hold $\P_c>T/2$; cf.~(\ref{krug}). \\ (b) Gray curve: the summary pay-off
$\P+\F$ given by (\ref{b1}, \ref{b2}) for same parameters as in (a), but
$\Delta=\frac{T}{2}-P$.  The maximal value of $\P+\F$ equals $T$. For
this specific choice of $\Delta$, $\I$ and $\II$ get equal pay-off:
$\P=\F=T/2$; see (\ref{b1}, \ref{b2}).} 
\label{figtwo}
\end{figure*}

\begin{figure}[!ht] \centering
{\includegraphics[width=8cm]{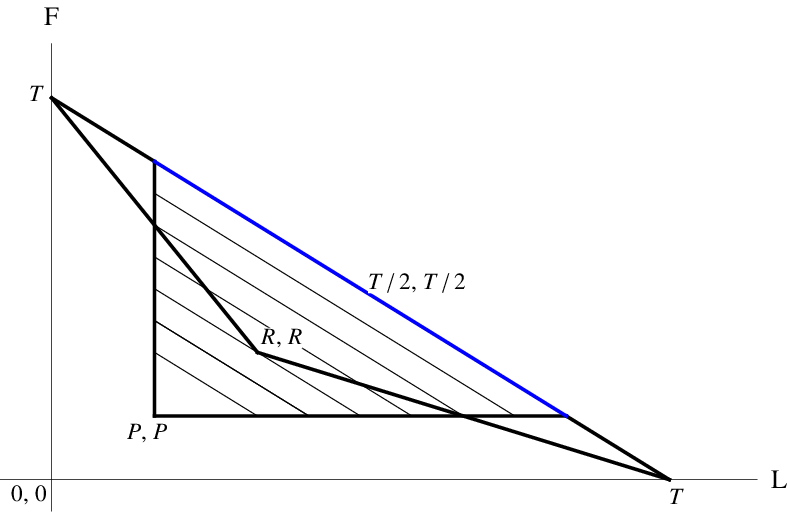}} 
\caption{ The same as in Fig.~\ref{fig2}, but with 
$T=3, R=1, P=\frac{1}{2}$, i.e. in the regime (\ref{rashid}). All pay-offs on the
blue line are achieved under certain values of $\Delta$ and $x$ related via (\ref{dr}).}
\label{fig3}
\end{figure}

Let us now apply (\ref{ukht}--\ref{b2}) to the case where 
\be
\no{rashid}
T>2=2R,
\ee
in (\ref{bad}). This situation is frequently omitted from the
consideration of the prisoner's dilemma game, but there are straightforward
examples, where it can be realized \footnote{Consider two firms ($\I$
and $\II$) that sell similar goods. Actions $c$ ($d$)
refer to advertising (not advertising) the goods. If both do not
advertise, then they share the market and sell equal amounts. If both
advertise, then they mutually neutralize each other, again share the
market, but now waste resources for advertising. However, if one
advertises and another does not, the one that advertised can take the
lion's share of the market. Now condition (\ref{rashid}) can be
realized in that situation. Moreover, (\ref{disu2}) is a type of cartel agreement,
where $\I$ does advertise, if $\II$ did not, and {\it vice versa}.}.
We note that under (\ref{rashid}), function $\P(x)$ locally minimizes at
$x=1$, and maximizes at $p_2^*=1$, i.e. whenever two arguments of ${\rm
max}[...,...]$ in (\ref{b3}) are equal; see Figs.~\ref{figtwo} for illustration. 
This leads to the value of $x=x_{\rm max}$
that maximizes $\P$ for a fixed $\Delta$:
\be
\label{dr}
x_{\rm max}=1-({P+\Delta})/{T},
\ee
and which automatically holds (\ref{cau}). Then we get from (\ref{b1}, \ref{b2}, \ref{dr}):
\be
\label{brr}
&& \P_{\rm max}\equiv\P(x_{\rm max})=T-P-\Delta=T-\F, \\
&& p^*_1(x_{\rm max})=0, \qquad p^*_2(x_{\rm max})=1.
\label{disu2}
\ee
Note from (\ref{brr}) that the summary (average) pay-off is now equal to its
maximal value. In this context, recall from the prisoner's dilemma game
(\ref{bad}) under condition (\ref{rashid}) that the deterministic
one-shot game has the maximal summary pay-off $2R=2$, but the average
maximal summary pay-off is $T>2R=2$. Next, (\ref{disu2}) shows that the
solution is in a sense fair, since $\I$ responds by $c$ to $d$, and by
$d$ to $c$; cf. (\ref{de}). Indeed, within (\ref{brr}) the pay-off of
$\II$ need not be smaller than that of $\I$. In particular, we gets
equal pay-offs $\P=\F=T/2$ for $x=1/2$.  More generally, besides holding
(\ref{cau}) and $x_{\rm max}<1$, $\Delta$ should be such that $\P_{\rm
max}>P$, which we shall assume to be the case. As seen in
Fig.~\ref{fig3}, $\P_{\rm max}$ and $\F$ fill the Pareto line on the
pay-off diagram in the regime $T>2=2R$. 

We now discuss how the transition from $x=1$ to $x<1$ can take place.
The difference between pay-offs reads from 
(\ref{brr}, \ref{b2}, \ref{shan}, \ref{caroussel}):
\be
\no{ga0}
\P_{\rm max}-\P_c &=& P(T-2)+\delta(T-1-\frac{\Delta}{\delta}), \\
\F-\F_c &=& \Delta-\delta.
\no{ga3}
\ee
It is seen that for $T>2$ there are situations, for a sufficiently small $\frac{\Delta}{\delta}>1$,
where both $\P_{\rm max}>\P_c$ and $\F>\F_c$ can hold, i.e. both $\I$ and $\II$ can benefit 
out of allowing $\II$ to defect. For $T<2$ such situations are excluded, as seen from (\ref{ga0}). 
Figs.~\ref{figtwo} demonstrate the behavior of $\P(x)$ for pertinent values of $x$.

We return to (\ref{de}--\ref{caroussel}) and note that they can
dissatisfy $\II$ due to two reasons: (a) $\II$ is not supposed to defect
due to ${\rm Pr}[\I\mapsto d\,|\, \II\mapsto d\,]=1$; cf.~(\ref{de}).
(b) His pay-off is small: $\P_c>\F_c$.  In response, $\I$ can propose
the set-up (\ref{dud}) that|though being more complex|is beneficial for
both for the assumed $T>2$ case. Fig.~\ref{fig} shows that there is a
range of $x\in (x_1,x_2)$, where both $\P>\P_c$ and $\F>\F_c$ hold.
(Obviously, $\P-\P_c>0$ cannot hold for $x=0$ or $x=1$.) For $x\in
(x_1,x_2)$, $p_2$ is well above zero, i.e. defections of $\II$ are
tolerated. Hence $x\in (x_1,x_2)$ is a regime that $\II$ can accept: he
is not punished for defection, and his pay-off increased compared to
$x=1$. $\I$ can offer this regime to $\II$, since his pay-off also
increases: $\P>\P_c$. 

Note that a conscientious $\II$ may wish to accept ${\rm Pr}[\I\mapsto
d\,|\, \II\mapsto d\,]=1$ as something rightful, but would like to
increase $p_1={\rm Pr}[\I\mapsto c\,|\, \II\mapsto c\,]$ from the value
$P+\delta$ (at $x=1$); see (\ref{de}--\ref{caroussel}). Now using 
(\ref{ukht}, \ref{b1}, \ref{b3}) we obtain
\be
\P-\P_c &=& -(T-1)(\Delta-\delta)+T(1-x[(T-P-1)p_2^* +P-1])\nonumber\\
&=& -(T-1)x(\,p_1^*-[P+\delta]\,)+(1-x)\left[\, (T-1)\delta+T(P-1)-p_2^* P\,\right],
\no{gaa2}
\ee
where $(T-1)\delta+T(P-1)<0$ due to (\ref{de}). Hence (\ref{gaa2}) show that 
$p_1>P+\delta$ always (i.e. both for $T<2R$ and $T>2R$) leads to $\P<\P_c$.
An example of this is seen in Fig.~\ref{fig}, where $\P(x)<\P_c$ whenever 
$p_1>P+\delta$.

\comment{
In sum, for $T>2$, $\I$ can increase his own pay-off by allowing
$\II$ to defect with probability $x\in (x_1,x_2)$.  For a particular
case of $x=x_{\rm max}$ the summary pay-off can reach its maximal value
$T$.}

\section{Discussion: Open problems of leadership}
\label{disco}

\comment{
To summarize the presented probabilistic commitment solution to the
prisoner's dilemma, we emphasize that it is based on the probabilistic
response of the leader $\I$ to the cooperation by the follower $\II$.
Hence a peculiar point of this solution is that $\I$ defects with a
certain probability; cf.~(\ref{de}). This makes the situation asymmetric
with respect to $\I$ and $\II$, i.e. this is a real leadership scenario
here. This asymmetry leads to a larger stability domain of the solution.
We recall that the autocratic (Stackelberg's) leadership leads to the
inefficient (trivial) solution, where both $\I$ and $\II$ defect each
other. Put differently, the autocratic leader is not powerful in several
important situations.  Recall that the power of the leader over the
follower is defined by the extent $\I$ can get $\II$ to do
things that $\II$ would otherwise not do \cite{riker}. Now the
autocratic leader is not powerful, e.g. because he will never change the
behavior of the follower who has a dominant strategy, i.e. a strategy
that is the best-response to all strategies of the leader. 

Thus the considered solution is a leadership scenario that is able to
resolve the prisoner's dilemma by transforming it such that $\II$ is not
motivated to defect, and $\I$ is motivated to keep his promise of egalitarian
leadership also due to a control from $\II$. 
}

Our results provide model-dependent answers to several
open problems of the leadership research. We reproduce these
questions from literature keeping their original form whenever possible.

{\bf 1:}
{\it Not do leaders make a difference, but under what conditions
does leadership matter \cite{hackman}? }

The above leadership scenario provides a solution to the prisoner's
dilemma, i.e. the leadership in the model has a functional role. The
leader ($\I$) here does have a power over the follower ($\II$), since he
convinced $\II$ not play the dominant strategy.  Recall that the power
of $\I$ over $\II$ is defined by the extent $\I$ can get $\II$ to do
things that $\II$ would otherwise not do \cite{riker}. The autocratic
(coercive or Stackelberg's) leader is not powerful in the (single-shot)
prisoner's dilemma game, because he will never change the behavior of the
follower who has a dominant strategy, i.e. a strategy that is the
best-response to all strategies of the leader. 

{\bf 2:}
{\it Are good and poor leadership qualitatively different
phenomena \cite{hackman}? }

The model shows a {\it qualitative} difference between good
(=egalitarian) vs.  poor (=exploiting) leadership: the summary pay-off|that
determines competitive abilities of the group as a whole|is higher or
equal for the egalitarian leadership. 

{\bf 3:}
{\it Under what conditions leaders exploit followers \cite{king,jtb,jtb2,evolution}? }

Within this model, exploitation|where the dilemma is resolved, but $\I$
gets more than $\II$|has certain rational roots, because the
exploitation regime has a larger stability domain in the sense explained
in sections \ref{condo} and \ref{stability}. Within the repeated-game
implementation of the scenario, the stability relates to back-reaction
of $\II$ on $\I$, and this does imply costs for $\II$. Another stability
mechanism (within the one-shot implementation) relates to the prestige
(reputation) of $\II$. 

{\bf 4:} {\it What are game-theoretic formalizations of the
Machiavellianism \cite{mac1}? }

The behavior of $\I$ does resemble the Machiavellian personality
extensively discussed in literature \cite{mac1,mac2}. This personality
is defined by three features: the ability to manipulate and exploit
people without provoking them. The Machiavellian personality is
considered as a possible model for social intellect \cite{mac1,mac2}.
First of all, we note that almost all distinguishing features of
manipulation \cite{anne,rudinow,dotsenko}|where $\I$ and $\II$ are
(resp.) the manipulator and manipulee|are present with the above
leadership scenario:

(1.) The manipulee does keep the full freedom of will. Indeed, in our
model $\II$ does have the right of the first move. There is no coercion
on $\II$. 

(2.) $\II$ need not be directly deceived \cite{rudinow}. In our model
$\II$ can know beforehand about all the pay-offs and the rules of the
game, which will be followed strictly by $\I$. 

(3.) The manipulator lets the manipulee to succumb to a weakness
\cite{anne}. In our model $\I$ can legitimize the exploitation by
arguing that it is more stable than the egalitarian regime; see the
discussion in sections \ref{condo} and \ref{stability}. $\II$ may agree
with this being afraid to loose his above-guaranteed pay-off (the
weakness). 

One of defining features of the Machiavellian personality
is also present in the model: 

(4.) Not provoking (skillfulness): $\I$ is ready to accept some limited
control from $\II$; see section \ref{stability}. In a different scenario
(realized for $T>2R$): if $\II$ protests against defections of $\I$, the
latter can re-manipulate the situation and permit $\II$ to defect with a
well-chosen probability thereby increasing his own pay-off; see section
\ref{preto}. 

Hence the exploiting leadership can be a model for Machiavellian
personality, which so far was not properly formalized within game
theory. Ref.~\cite{mac1} directly related this personality with the
defection strategy of the prisoner's dilemma game, but this relation does not
account for its manipulative aspects and hence does not explain why it
is a form of social intellect. The defection strategy of the prisoner's dilemma
game is easily neutralized by defecting in response.

{\bf 5:} {\it What are implications of the Machiavellian
leadership for followers \cite{evolution}? }

Since $\II$ is subject to manipulation and exploitation, he has to be
non-Machiavellian. Such people score low in corresponding tests
\cite{mac1,mac2}. Note that the back-reaction of $\II$ on $\I$ that
involves sequential defections of $\I$, also roughly coincides with
the behavior of non-Machiavellian followers. 

We studied two implications of the exploitative leadership for $\II$.
First, $\II$ can fall into apathy, because he is put into a situation,
where either action by him gives comparable average pay-offs. Another
option is that $\II$ will be actively dissatisfied by rules of the
game, i.e. by the fact that he gets less than $\I$ and he is
effectively prevented from defecting, while $\I$ does defect.  
Now $\I$ can solve both these problems simultaneously|and yet
to increase his own average pay-off|by allowing $\II$ to defect with a
certain (well-chosen) probability. This demands that the pay-off for
defecting a cooperating opponent is sufficiently large; see section
\ref{preto}. 

There is an evidence that dyadic groups composed of one
Machiavellian and one non-Machiavellian are out-performed by two
Machiavellians, or two non-Machiavellians that are able to develop
more egalitarian coalitions \cite{mac1}; cf. {\bf 2}. 

\comment{
It is not widely recognized that the Machiavellian personality is a form 
of leadership. In its non-trivial forms it emerges out of solving
certain problems. We show that the prisoner's dilemma solving leadership
can develop into manipulation due to objective reasons, i.e. th
follower $\II$ having a limited control over the leader $\I$. }

{\bf 6:} {\it Not what are the traits of leaders, but how do
leaders' personal attributes interact with situational properties to
shape outcomes \cite{hackman}? }

In the studied game-theoretic model, $\I$ does have personal attributes.
For $T<2R$, the leader $\I$ has a manipulative (Machiavellian
\cite{mac1,mac2}) social intellect that allows him to gain an additional
pay-off via probabilistic response, but $\I$ is also willing to admit a
limited control from $\II$ (via the repeated-game implementation), or
from his environment via prestige (reputation) factors. The final
pay-off of $\I$ is determined by the inter-play between these
mechanisms. Now these attributes are optimal (i.e. emerge out of pay-off
maximization) for the prisoner's dilemma game considered. For other symmetric
games they are inferior: the prisoner's dilemma (\ref{bad}) is the only
symmetric game that admits a non-trivial probabilistic response
(\ref{de}) \cite{sano}. E.g. for symmetric coordination games a
probabilistic response can be introduced, but it is inferior to purely
deterministic responses \cite{sano}. 

For $T>2R$ (see section \ref{control}), the leader $\I$ should deal with
the complexity of the situation, since $\I$ should to some extent
understand motivations of the follower $\II$ (i.e. why $\II$ is
defecting). Then it is possible to improve both pay-offs of $\I$ and
$\II$. 

\comment{
Attributes of the follower $\II$ are seen when looking at his reactions
against an over-exploiting $\I$. A trustless $\II$ does not care about
$\I$ defecting to his cooperation, but wants to have a right of
defecting without punishment, and wants to improve his pay-off. Then
$\I$ can allow $\II$ to defect with a certain controllable probability,
and both will gain from this. This controllable defection scenario does
not work with a conscientious $\II$ that will be insist on a larger
pay-off together with $\I$ defecting less to his cooperation. }

\comment{
{\bf 7:} {\it Not how do leaders and followers differ, but how can
leadership models be reframed so they treat all system members as both
leaders and followers \cite{hackman}? }

For the present model the traditional Stackelberg leadership|where $\I$
acts and $\II$ reacts|is not efficient. Instead, an inverse scenario is
efficient, where the right of the first move belongs to $\II$. Thus what
appears to be the main ingredient of the traditional leadership (the
first move) is now delegated to $\II$. 
}

{\bf 7:} {\it What can be said about the situation with many followers and more than one leader \cite{armen}? }

This interesting problem should be studied in future possibly via tools of statistical mechanics that were
recently applied to game theory \cite{szabo,adami,colin}. 

\section*{Acknowledgment} We thank A. Khachatryan for useful
discussions. AEA and SGB were supported by SCS of Armenia, grants No. 18RF-002 and No. 18T-1C090.
AVM was supported by RFBR via the research project 18-51-05007 arm\_\,a.

\appendix

\section{Extended representation of the probabilistic commitment}
\label{cons}

Let us write a generic $2\times 2$ game as 
\begin{eqnarray}
\begin{tabular}{||c|c|c||}
  \hline
~${\I}/{\II}$~  & $\alpha$ & $\beta$ \\
  \hline
  ~$i$~ & ~$s_{i\alpha}, \, \sigma_{i\alpha}$~ & ~$s_{i\beta}, \, \sigma_{i\beta}$~ \\
  \hline\hline
  ~$j$~ & ~$s_{j\alpha}, \, \sigma_{j\alpha}$~ & ~$s_{j\beta}, \,
  \sigma_{j\beta}$~ \\
  \hline
\end{tabular} ,
\label{dadi}
\end{eqnarray}
where notations obviously generalize (\ref{bad}).

Let now the game (\ref{dadi}) be considered in the sequential form:
$\II$ makes the first move with the strategies $\a$ or $\b$, while the
strategies of $\I$ are announced beforehands, i.e. before knowing the
move of $\II$. Hence there are four of them: $ii$ and $jj$ mean that
$\I$ will act (resp.) $i$ and $j$ for any move of $\II$, $ij$ means that
$\I$ will act $i$ ($j$) in response to $\a$ ($\b$), while $ji$ means
that $\I$ will act $j$ ($i$) in response to $\a$ ($\b$). 
We represent the situation as follows: 
\begin{eqnarray}
\begin{tabular}{||c|c|c||}
  \hline
~${\I}/{\II}$~  & $\alpha$ & $\beta$ \\
  \hline
  ~$ii$~ & ~$s_{i\alpha}, \, \sigma_{i\alpha}$~ & ~$s_{i\beta}, \, \sigma_{i\beta}$~ \\
  \hline\hline
  ~$ij$~ & ~$s_{i\alpha}, \, \sigma_{i\alpha}$~ & ~$s_{j\beta}, \,   \sigma_{j\beta}$~ \\
  \hline\hline
  ~$ji$~ & ~$s_{j\alpha}, \, \sigma_{j\alpha}$~ & ~$s_{i\beta}, \, \sigma_{i\beta}$~ \\
  \hline\hline
  ~$jj$~ & ~$s_{j\alpha}, \, \sigma_{j\alpha}$~ & ~$s_{j\beta}, \,   \sigma_{j\beta}$~ \\
  \hline
\end{tabular}
\label{daad}
\end{eqnarray}

Let us now endow $\I$ and $\II$ with mixed strategies: the four
strategies of $\I$ get probabilities $P_{ii}$, $P_{ij}$, $P_{ji}$, and
$P_{jj}$ with $P_{ii}+P_{ij}+P_{ji}+P_{jj}=1$. Note that out of these
quantities we can construct conditional probabilities:
\be
P_{ii}+P_{ij}=P_{i|\a}, \quad
P_{jj}+P_{ji}=P_{j|\a}, \quad
P_{ii}+P_{ji}=P_{i|\b}, \quad
P_{jj}+P_{ij}=P_{j|\b}.
\label{mimi}
\ee
The two strategies of
$\II$ have probabilities $\pi_\a$ and $\pi_\b$ with $\pi_\a+\pi_\b=1$. 

Employing the prisoner's dilemma game pay-offs (\ref{bad}) in (\ref{daad}), we
get $(d,dd)$ is the only Nash equilibrium of the game. Putting into
(\ref{dadi}) pay-offs of the ultimatum game as defined in section
\ref{ulti} we obtain:
\begin{eqnarray}
\begin{tabular}{||c|c|c||}
  \hline
~${II}/{I}$~  & $f$ & $u$ \\
  \hline
  ~$a$~ & ~$(0.5\,\mathfrak{M}, \, 0.5\,\mathfrak{M})$~ & ~$(0.01\,\mathfrak{M}, \,  0.99\,\mathfrak{M} )$~ \\
  \hline\hline
  ~$r$~ & ~$(0\,,0 )$~ & ~$(0\,,0 )$  ~ \\
  \hline
\end{tabular}\, ,
\label{ultii}
\end{eqnarray}
where $a$ and $r$ ($f$ and $u$) are actions of $II$ ($I$). Recall that
$II$ makes the second move.  There is a single Nash equilibrium in
(\ref{ultii}): $(u,a)$. Now using (\ref{ultii}) in (\ref{daad}), we
shall get two equilibria: $(u,aa)$ (that resembles the previous one), and
$(f, ar)$, where $II$ treatens $I$ to reject the unfair offer. This
second Nash equilibrium is not sub-game perfect (credible), since $r$ is
never the best response of $II$. 

\comment{
Let us write standard conditions of the (mixed) Nash equilibrium at $\{ P_{i|\a}, P_{i|\b}; \pi_a\}$ for the game
(\ref{daad}):
\be
\label{go}
P_{i|\a}s_{i\a}\pi_\a + P_{j|\a}s_{j\a}\pi_\a + P_{i|\b}s_{i\b}\pi_\b + P_{j|\b} s_{j\b}\pi_\b\geq 
P'_{i|\a}s_{i\a}\pi_\a + P'_{j|\a}s_{j\a}\pi_\a + P'_{i|\b}s_{i\b}\pi_\b + P'_{j|\b} s_{j\b}\pi_\b,\\
P_{i|\a}\sigma_{i\a}\pi_\a + P_{j|\a}\sigma_{j\a}\pi_\a + P_{i|\b}\sigma_{i\b}\pi_\b + P_{j|\b} \sigma_{j\b}\pi_\b\geq 
P_{i|\a}\sigma_{i\a}\pi'_\a + P_{j|\a}\sigma_{j\a}\pi'_\a + P_{i|\b}\sigma_{i\b}\pi'_\b + P_{j|\b} \sigma_{j\b}\pi'_\b,
\label{ra}
\ee
where $\{ P'_{i|\a}, P'_{i|\b}; \pi'_a\}$ are any other probabilities.
The message of (\ref{go}) [(\ref{ra})] is well-known: any unilaterial
deviation of $\I$ from $\{ P_{i|\a}, P_{i|\b}\}$ [of $\II$ from
$\{\pi_a\}$] cannot increase their average pay-offs. 
}

\section{Second-order metagames}
\label{meta2}

Here we shall present the second-order metagame approach
\cite{howard,shubik} in a slightly generalized, probabilistic form that
will make obvious its main assumption. Given the game (\ref{dadi}), let
us define conditional probabilities $P^{[a]}_{i|\a}=1-P^{[a]}_{j|\a}$
and $P^{[a]}_{i|\b}=1-P^{[a]}_{j|\b}$ for the response of $\I$ to pure
strategies of $\II$.  The index $a=1,2,..,K_\I$ refers to different
possible responses (i.e. meta-strategies of $\I$), their number and form
need not be specified for our present purposes.  Then the first-order
metagame is defined by average pay-offs:
\be
\label{ord}
\widetilde{s}_{a\a}\equiv P^{[a]}_{i|\a}s_{i\a} + P^{[a]}_{j|\a}s_{j\a}, 
\qquad
\widetilde{\sigma}_{a\a}\equiv P^{[a]}_{i|\a}\sigma_{i\a}+ P^{[a]}_{j|\a}\sigma_{j\a},
\ee
for $\I$ and $\II$, respectively. The meaning of (\ref{ord})
is clear and refers to averages over pure strategies of $\I$,
given the probabilistic response of $\I$
to pure strategies of $\II$.

Starting from (\ref{ord}), Howard \cite{howard} (see \cite{shubik} for a
review) proceeds to define the second-order meta-game approach, where
$\II$ now probabilistically reacts to meta-strategies of $\I$. These
reactions are described by conditional probabilities $P^{[b]}_{\a|a}$,
where $b=1,2,..,K_\II$. Although Howard restricted the situation to pure
(non-mixed) responses $P^{[b]}_{\a|a}=0,1$, this restriction is not
essential for the argument we want to make. 

This invites to define the second-order pay-offs by analogy to (\ref{ord}):
\be
\label{orda}
\widetilde{\widetilde{s}}_{ab}\equiv \widetilde{s}_{a\a} P^{[b]}_{\a|a}+\widetilde{s}_{a\b} P^{[b]}_{\b|a},
\qquad
\widetilde{\widetilde{\sigma}}_{ab}\equiv \widetilde{\sigma}_{a\a} P^{[b]}_{\a|a}+ \widetilde{\sigma}_{a\b} P^{[b]}_{\b|a}.
\ee
Separate terms in (\ref{orda}) are interpreted as follows; e.g.
$s_{i\a}P^{[a]}_{i|\a} P^{[b]}_{\a|a}$ that enters into the first
equation in (\ref{orda}) means that $\I$ first announces his reaction
$a$ (commits to $a$), then $\II$ acts in response $\a$ with probability
$P^{[b]}_{\a|a}$, and then $\I$ acts $i$ with the promised reaction
(probability) $P^{[a]}_{i|\a}$. Thus the strategies of $\I$ ($\II$)
amount to $a=1,2,..,K_\I$ ($b=1,2,..,K_\II$). It is important to stress
here that the consistency of the last move of $\I$ with its own
commitment is postulated here. The credibility (i.e. stability) of such
a commitment was studied in \cite{reno}.

\end{document}